\documentclass[final,3p,times,twocolumn]{elsarticle}

\usepackage{blindtext, graphicx, amsmath, algorithm, algpseudocode, pifont, algcompatible, comment, layout, amsthm, amssymb}
\usepackage{enumitem}   
\usepackage{eso-pic}
\usepackage{booktabs}
\usepackage{float}
\usepackage{ulem}
\usepackage{xcolor}
\usepackage{listings}
\usepackage[T1]{fontenc}
\usepackage{tikz}
\usepackage{pgfplots}
\pgfplotsset{compat=newest}
\usepgfplotslibrary{groupplots}

\usepackage{kotex}
\usepackage{kotex-logo}

\usepackage[utf8]{inputenc}
\usepackage[english]{babel}
\usepackage{hyperref} 
\hypersetup{ colorlinks=true, linkcolor=black, filecolor=black, urlcolor=cyan, }
\usepackage{adjustbox}
\usepackage{caption}
\journal{ICT Express}

\begin{document}

\begin{frontmatter}

\title{BemaGANv2: Discriminator Combination Strategies for GAN-based Vocoders in Long-Term Audio Generation}
\author[1]{Taesoo Park}
\ead{taesoo0707@kw.ac.kr}

\author[1]{Mungwi Jeong}
\ead{ansrnl19178@kw.ac.kr}

\author[3]{Mingyu Park}
\ead{mingyu.park@dgist.ac.kr}

\author[1]{Narae Kim}
\ead{wing02@kw.ac.kr}

\author[1]{Junyoung Kim}
\ead{kimjunyoung@kw.ac.kr}

\author[1]{Mujung Kim}
\ead{kmj1026@kw.ac.kr}

\author[1]{Jisang Yoo}
\ead{jsyoo@kw.ac.kr}

\author[4]{Hoyun Lee}
\ead{hoyun1004@gmail.com}

\author[5]{Sanghoon Kim}
\ead{hoon0700@naver.com}

\author[2]{Soonchul Kwon\corref{cor1}}
\ead{ksc0226@kw.ac.kr}

\address[1]{Department of Electronic Engineering, Kwangwoon University, Seoul, South Korea}
\address[2]{Graduate School of Smart Convergence, Kwangwoon University, Seoul, South Korea}
\address[3]{Department of Electrical and Electronic Computer Engineering, DGIST, Daegu, South Korea}
\address[4]{Ewha Womans University College of Medicine Seoul, South Korea}
\address[5]{School of Medicine, Kyung Hee University Seoul, South Korea}

\cortext[cor1]{Corresponding author}

\begin{abstract}
This paper presents BemaGANv2, an advanced GAN-based vocoder designed for high-fidelity and long-term audio generation, with a focus on systematic evaluation of discriminator combination strategies. Long-term audio generation is critical for applications in Text-to-Music (TTM) and Text-to-Audio (TTA) systems, where maintaining temporal coherence, prosodic consistency, and harmonic structure over extended durations remains a significant challenge.
Built upon the original BemaGAN architecture, BemaGANv2 incorporates major architectural innovations by replacing traditional ResBlocks in the generator with the Anti-aliased Multi-Periodicity composition (AMP) module, which internally applies the Snake activation function to better model periodic structures. 
In the discriminator framework, we integrate the Multi-Envelope Discriminator (MED), a novel architecture we proposed, to extract rich temporal envelope features crucial for periodicity detection. 
Coupled with the Multi-Resolution Discriminator (MRD), this combination enables more accurate modeling of long-range dependencies in audio. 
We systematically evaluate various discriminator configurations, including Multi-Scale Discriminator (MSD) + MED, MSD + MRD, and Multi-Period Discriminator (MPD) + MED + MRD, using objective metrics (Fréchet Audio Distance (FAD), Structural Similarity Index (SSIM), Pearson Correlation Coefficient (PCC), Mel-Cepstral Distortion (MCD), Multi-Resolution STFT (M-STFT), Periodicity error (Periodicity)) and subjective evaluations (MOS, SMOS). 
To support reproducibility, we provide detailed architectural descriptions, training configurations, and complete implementation details. The code, pre-trained models, and audio demo samples are available at: \url{https://github.com/dinhoitt/BemaGANv2}.
\end{abstract}


\end{frontmatter}


\section{Introduction}\label{sec1}

In recent years, deep learning-based generative models for Text-to-Audio (TTA) and Text-to-Music (TTM) have gained significant attention due to their capability to produce high-fidelity audio. In diffusion-based architectures, the vocoder serves as a critical component that transforms intermediate acoustic representations such as Mel-spectrograms into time-domain waveform signals. The quality and structure of the vocoder directly influence the realism and continuity of the generated audio, particularly in long-form generation tasks.

Early models such as MelGAN~\cite{b1} emphasized inference efficiency through fully convolutional architectures but struggled with harmonic fidelity. HiFi-GAN~\cite{b2} addressed this by introducing a Multi-Period Discriminator (MPD) that improved periodicity modeling. Subsequently, BigVGAN extended this line of work with the Anti-aliased Multi-Periodicity (AMP) module and the Snake activation~\cite{b3, b4} to better capture harmonic structures and long-term coherence.

Our previous work, BemaGAN~\cite{b5}, introduced the Multi-Envelope Discriminator (MED), which focuses on capturing temporal energy envelopes through both Hilbert transform-based envelope extraction (for upper and lower bounds) and low-pass filtered signals (for multi-scale amplitude variations ). Building on this, BemaGANv2 combines several architectural enhancements: the AMP block replaces the ResBlock in the generator; the Snake activation is employed to improve periodic signal modeling; and the discriminator is extended with the Multi-Resolution Discriminator (MRD) to refine spectral consistency~\cite{b6}. Detailed descriptions of the AMP block, Snake activation, and MRD are provided in Section~\ref{sec3}. While diffusion-based vocoders such as WaveGrad~\cite{b7} and DiffWave~\cite{b8} have demonstrated impressive audio quality, their iterative sampling process increases computational cost and latency, making them less suitable for long-term generation and real-time streaming applications. Therefore, this work focuses exclusively on GAN-based vocoders, which offer fast, single-pass inference while maintaining high fidelity—a critical requirement for deployment in TTA and TTM systems.

This paper provides a tutorial-style study of GAN-based vocoders with a focus on discriminator design for long-term audio generation. While diffusion-based, autoregressive, and flow-based vocoders represent important parallel research directions, we limit our scope to GAN-based models to maintain coherence and alignment with long-term generation and real-time streaming requirements. Specifically, our goals are threefold:
\begin{itemize}
    \item To provide a tutorial-style overview of the core architectural principles, training methodology, and implementation strategies of modern GAN-based vocoders.
    \item To systematically analyze how different discriminator combinations affect audio reconstruction quality under identical generator and training conditions, demonstrating that the choice of discriminator combination is a critical design factor.
    \item To present BemaGANv2 (MED+MRD) as a case study demonstrating that complementary discriminator pairing—temporal envelope modeling via MED and spectral consistency via MRD—achieves the most balanced performance, rather than simply replacing any single existing discriminator.
\end{itemize}

All experiments discussed herein were conducted using the LJSpeech dataset for training and freeform audio from Freesound.org for evaluation. Computation was performed on an NVIDIA A100 GPU via Google Colab.

LJSpeech Dataset \cite{b20}: A widely-used single-speaker English speech corpus consisting of 13,100 short audio clips with a total duration of approximately 24 hours. Recorded by professional speaker Linda Johnson, the dataset features clean, high-quality 16-bit PCM audio at 22,050 Hz sampling rate. LJSpeech has become a standard benchmark for speech synthesis and vocoder evaluation due to its consistent recording quality and phonetic diversity.

Freesound.org Audio: A collaborative database of Creative Commons-licensed audio samples spanning diverse acoustic conditions, including environmental sounds, musical instruments, sound effects, and ambient recordings. We selected samples from Freesound.org to evaluate model generalization beyond speech, testing vocoder performance on out-of-distribution audio with complex spectral characteristics and long-term temporal structures.

This paper is an extended version of our prior work published at ICAIIC 2025~\cite{b6}, enriched with additional experiments, comparative discussions, and theoretical analysis. The content has been restructured to support reproducibility, enriched with ablation experiments, expanded evaluation metrics, and theoretical analysis.

\section{Related Work} \label{sec2}

\subsection{Overview of Vocoders} \label{sec2.1}

In TTA and TTM systems, the vocoder plays a critical role in converting intermediate acoustic representations—typically Mel-spectrograms—into time-domain audio waveforms. The quality and structure of the vocoder significantly affect the fidelity, naturalness, and temporal coherence of the generated audio. In fact, most diffusion-based TTA and TTM models, such as AudioLDM\cite{b9,b10},FLUX Music\cite{b11}, commonly adopt HiFi-GAN as the vocoder due to its high-quality reconstruction and efficient inference. In this subsection, we review key developments in parallel GAN-based vocoders, focusing on MelGAN, HiFi-GAN, and BigVGAN, to illustrate the evolution of architecture design and its implications for periodicity modeling. Following this, we also discuss our prior work, BemaGAN , and its advanced iteration, BemaGANv2, which serve as a culmination of these architectural advancements.

\subsubsection{MelGAN} \label{sec2.1.1}
MelGAN was one of the earliest successful attempts to apply generative adversarial networks (GANs) to the task of non-autoregressive raw waveform generation. The model adopts a fully convolutional generator that transforms Mel-spectrograms into audio waveforms in a feed-forward and parallel manner, enabling real-time synthesis on both GPU and CPU without the need for autoregressive decoding or teacher-student distillation.

The generator in MelGAN employs transposed convolutional layers for upsampling, each followed by residual stacks with dilated convolutions to expand the receptive field and capture long-range dependencies. The design also avoids using global noise vectors, relying solely on the conditioning Mel-spectrogram as input. This was motivated by empirical findings that strong conditioning signals are sufficient to generate coherent audio without additional stochasticity.

For the discriminator, MelGAN introduces a multi-scale discriminator (MSD) architecture, where three sub-discriminators operate on different temporal resolutions of the input waveform. This design allows the discriminator to capture audio structures at varying scales, from local textures to global continuity\cite{b1}.

Although MelGAN demonstrated impressive speed and efficiency—with more than 2500 kHz inference speed on a single GPU and a small parameter footprint (~4.2M)—it showed perceptual quality limitations compared to autoregressive models like WaveNet or flow-based models like WaveGlow. Later studies, including HiFi-GAN, extended this line of work improved fidelity through enhanced discriminator designs and loss functions\cite{b2}.

\subsubsection{HiFi-GAN} \label{sec2.1.2}
HiFi-GAN is a high-fidelity and efficient GAN-based vocoder that significantly improved upon previous models like MelGAN by refining both the generator and the discriminator architectures. While it retains the Multi-Scale Discriminator (MSD) introduced in MelGAN, HiFi-GAN adds a novel Multi-Period Discriminator (MPD) to more effectively capture the periodic structures of waveform signals\cite{b2}.

The generator in HiFi-GAN adopts a stack of upsampling layers followed by multi-receptive field fusion (MRF) modules, which consist of multiple residual blocks with varying kernel sizes and dilation rates. This design enables the generator to capture both local and global patterns in the signal, contributing to better harmonic reconstruction\cite{b2}.

The Multi-Period Discriminator (MPD) is the key innovation in HiFi-GAN’s discriminator framework. Each sub-discriminator in the Multi-Period Discriminator (MPD) reshapes the 1-D waveform into 2-D matrices using different periodic lengths and applies 2-D convolutions to capture periodic patterns over varying cycles. This architecture complements the Multi-Scale Discriminator (MSD), which focuses on multi-scale representations in the time domain via downsampling\cite{b2}.

HiFi-GAN achieves high fidelity in waveform generation while maintaining extremely fast inference speed (e.g., 93.75× faster than real-time on GPU), making it suitable for real-time TTS systems. However, its reliance on Leaky ReLU activation functions and conventional residual blocks limits its capacity to learn fine-grained periodicity and harmonic richness in more complex or out-of-distribution (OOD) data, as pointed out in subsequent works like BigVGAN\cite{b3}.

\begin{table*}[t]
\centering
\caption{Comparison of GAN-based vocoder architectures.}
\label{tab:vocoder_comparison}
\resizebox{\textwidth}{!}{%
\begin{tabular}{@{}lccccc@{}}
\toprule
\textbf{Model} & \textbf{Generator} & \textbf{Discriminators} & \textbf{Activation} & \textbf{Key Features} & \textbf{Limitations} \\
\midrule
MelGAN & Transposed Conv + ResBlock & MSD & Leaky ReLU & Real-time inference & Poor periodicity \\
HiFi-GAN & MRF-based ResBlocks & MPD + MSD & Leaky ReLU & Harmonic modeling & Fixed periodic hops \\
BigVGAN & AMP blocks & MPD + MRD & Snake & Anti-aliasing, expressive periodicity & High complexity \\
BemaGAN & HiFi-GAN ResBlock & MPD + MED & Leaky ReLU & Envelope-aware detection & Generator underpowered \\
\textbf{BemaGANv2} & AMP block & MED + MRD & Snake & Balanced temporal and spectral modeling & Single-speaker training data \\
\bottomrule
\end{tabular}%
}
\end{table*}

As a result, HiFi-GAN has served as a strong baseline for later vocoders aiming to improve both periodicity modeling and generalization, including BigVGAN and the BemaGAN series.

\subsubsection{BigVGAN} \label{sec2.1.3}
BigVGAN is a universal GAN-based vocoder designed to synthesize high-fidelity audio across a wide range of conditions, including unseen speakers, languages, and noisy recording environments. It builds upon the generator architecture of HiFi-GAN but introduces several architectural and training innovations to address the limitations of prior models in capturing periodicity, extrapolation, and robustness to out-of-distribution (OOD) inputs\cite{b3, b4}.

The core architectural contribution of BigVGAN is the AMP (Anti-aliased Multi-Periodicity) block, which replaces the traditional residual blocks in the generator. Each AMP block embeds channel-wise Snake activation functions within a series of dilated convolutions, thereby enforcing an explicit periodic inductive bias in the transformation process.
 Each AMP block contains a series of dilated convolutions with channel-wise Snake activation functions, which provide an explicit periodic inductive bias. The Snake function, defined as:

\begin{equation} f_{\alpha}(x) = x + \frac{1}{\alpha} \sin^{2}(\alpha x)\label{eq1} \end{equation}

where α is a learnable frequency parameter that controls the periodicity 
of the activation. A higher α increases the frequency of oscillation, 
allowing the network to model fine-grained harmonic details, while a 
lower α captures broader periodic structures. In our implementation, α is 
initialized randomly and optimized during training via backpropagation.

This learnable periodicity significantly improves the model's ability to 
generalize and extrapolate periodic content, which is particularly 
beneficial in musical or expressive vocal signals\cite{b3}

To prevent aliasing artifacts caused by high-frequency nonlinearities, BigVGAN applies a low-pass filter to the output of the Snake activation via a process of anti-aliased upsampling and downsampling, inspired by alias-free design principles from image generation models like StyleGAN3\cite{b3,b12}.

In the discriminator, BigVGAN replaces HiFi-GAN’s Multi-Scale Discriminator (MSD) with a Multi-Resolution Discriminator (MRD), which operates on spectrograms of various STFT configurations. This Multi-Resolution Discriminator (MRD), originally proposed in UnivNet, enables enhanced spectral sharpness and pitch accuracy by leveraging multiple time-frequency resolutions\cite{b13}.

BigVGAN is also notable for its large-scale training: the full version of the model contains over 112 million parameters and is trained on the entire LibriTTS corpus, including diverse environments. Despite the increased capacity, it achieves faster-than-real-time inference and state-of-the-art results on both in-distribution and out-of-distribution tasks, outperforming HiFi-GAN and other vocoders in both MOS and SMOS metrics.

In summary, BigVGAN demonstrates that enhancing the generator’s inductive bias through periodic activations and anti-aliasing, along with robust discriminator design and scaling strategies, leads to significant improvements in universal vocoding performance.
However, the increased architectural complexity and sensitivity during training highlight the need for more efficient and stable vocoder designs in future work.

\subsection{Snake Function and Periodictiy in Audio} \label{sec2.2}

\begin{figure*}[t]
    \centering
    \begin{tikzpicture}
    \begin{groupplot}[
        group style={
            group size=3 by 2,
            horizontal sep=1.6cm,
            vertical sep=1.6cm,
        },
        width=0.30\textwidth,
        height=4cm,
        grid=both,
        xlabel={$x$},
        ylabel={$f(x)$},
    ]

    \nextgroupplot[title={ReLU}]
    \addplot[blue, thick] {max(0, x)};

    \nextgroupplot[title={Leaky ReLU}]
    \addplot[orange, thick] {x < 0 ? 0.1*x : x};

    \nextgroupplot[title={Tanh}]
    \addplot[red, thick] {tanh(x)};

    \nextgroupplot[title={Sin}]
    \addplot[green, thick] {sin(deg(x))};

    \nextgroupplot[title={Snake ($\alpha=1$)}]
    \addplot[purple, thick] {x + sin(deg(x))^2};

    \end{groupplot}
    \end{tikzpicture}
    \caption{Comparison of activation functions (each in separate axis).}
\end{figure*}

This limitation is closely tied to the inductive biases of common activation functions. Functions like ReLU and its variants extrapolate linearly, while tanh and sigmoid tend to saturate to constants. As a result, these activations are inherently ill-suited for learning or generating periodic signals, which require oscillatory behavior that persists outside the training interval.

To address this, recent research has proposed the Snake activation function, which introduces a trainable periodic component while maintaining monotonicity and optimization stability. Unlike earlier periodic activations such as sin or cos—which are difficult to optimize due to non-monotonicity and numerous local minima—Snake balances periodicity and trainability\cite{b4}.

Theoretically, the Snake activation satisfies a universal extrapolation property: neural networks using Snake can approximate any piecewise-smooth periodic function over the real line, not just within the training interval. Empirically, it has been shown to outperform ReLU, tanh, and sin in periodic regression tasks and time-series prediction.

In the context of neural vocoders, Snake provides a strong inductive bias for harmonic structure, enabling better modeling of voiced speech and musical tones. When embedded within advanced generator architectures, it enhances both periodicity modeling and signal continuity, making it a powerful component for high-fidelity audio synthesis.

\subsection{Architectural Comparisons and Motivations for BemaGANv2} \label{sec2.3}

While previous sections have detailed individual vocoder architectures, we here provide a comparative analysis focusing on their strengths, limitations, and how they motivated the design of BemaGANv2.

HiFi-GAN combines Multi-Scale Discriminator (MSD) and Multi-Period Discriminator (MPD) with a residual-based generator, enabling high-fidelity audio synthesis\cite{b2}. However, its generator lacks a strong inductive bias for periodicity, and the fixed-period hop sizes used in Multi-Period Discriminator (MPD) can be suboptimal for diverse or expressive signals.

BigVGAN improves upon this by introducing the AMP block with periodic activations (Snake) and replacing Multi-Scale Discriminator (MSD) with the Multi-Resolution Discriminator (MRD) for enhanced spectral sharpness\cite{b3}. However, this comes with a significant increase in parameter count and training complexity.

BemaGAN seeks to improve temporal dynamics modeling through the Multi-Envelope Discriminator (MED), but retains generator of HiFi-GAN, which limits the benefit of the improved discriminator\cite{b5}.

BemaGANv2 is thus designed to combine the periodicity modeling capability of the AMP module (used in BigVGAN) with the temporal envelope sensitivity of Multi-Envelope Discriminator (MED) and the spectral resolution of Multi-Resolution Discriminator (MRD). This MED+MRD combination enables better coverage of both time-domain and frequency-domain perceptual cues while keeping the model more lightweight and stable than BigVGAN.

This integration addresses the limitations of each predecessor and is empirically validated in Section~\ref{sec4}.
    
\section{Model} \label{sec3}

In this section, we provide a detailed tutorial-style description of the architectural components of BemaGANv2. The model adopts a GAN-based vocoder framework, integrating periodicity-aware generator and perceptually motivated discriminator. We begin by describing the generator based on AMP blocks, followed by the Multi-Envelope Discriminator (MED) and Multi-Resolution Discriminator (MRD), and conclude with the training objectives. This structured approach aims to provide a clear understanding of how each component contributes to BemaGANv2's overall performance in high-fidelity, long-term audio generation.

\subsection{Generator with AMP Block}

\begin{figure}[htbp]
\centering
\includegraphics[width=\columnwidth]{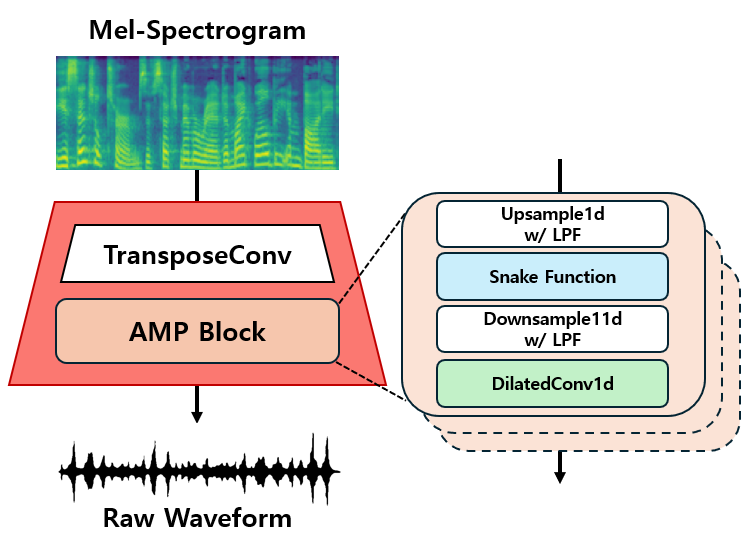}
\caption{The architecture of the AMP-based generator used in BemaGANv2. The AMP block, originally introduced in BigVGAN, integrates upsampling and downsampling operations with low-pass filtering (LPF), Snake activation for periodic inductive bias, and dilated convolutions.}
\label{fig1}
\end{figure}

The generator in BemaGANv2 is structured similarly to HiFi-GAN, utilizing transposed convolution layers for upsampling and residual blocks for waveform synthesis. However, a key distinction is its employment of the Anti-aliased Multi-Periodicity (AMP) block, originally introduced in BigVGAN\cite{b3}, which replaces conventional residual blocks.

Each AMP block, depicted in Figure~\ref{fig1}'s detailed view, consists of an upsampling layer with Low-Pass Filtering (LPF), followed by a Snake activation function, a downsampling layer with LPF, and dilated convolutions. The LPFs are crucial for anti-aliasing, preventing high-frequency artifacts that can arise from non-linear operations like upsampling and the Snake activation. Each AMP block contains dilated convolutions paired with the Snake activation function, which provides a learnable periodic bias. This allows the generator to better model harmonic structures and rhythmic patterns in long-term audio generation by injecting learnable periodicity into the activation, which is particularly beneficial in musical or expressive vocal signals. The overall design enhances both local detail and global periodic consistency.\cite{b3, b4}.

\subsection{Multi-Envelope Discriminator (MED)}

\begin{figure}[htbp]
\centering
\includegraphics[width=\columnwidth]{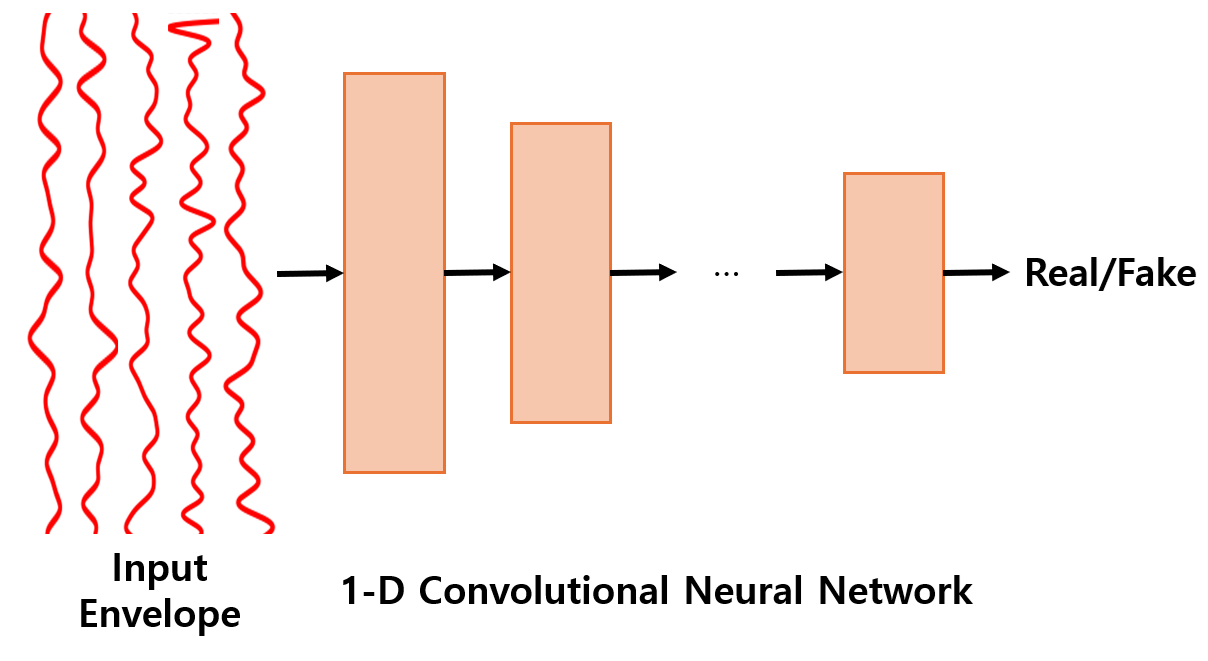}
\caption{ The structure of the Multi-Envelope Discriminator (MED). Time-domain envelopes, including both upper and lower envelopes, are extracted from the input audio using low-pass filters with different cutoff frequencies. These envelope signals are then processed by 1D convolutional layers. This design enables the discriminator to detect temporal energy patterns that are crucial for perceptual quality, such as prosodic variation.}
\label{fig2}
\end{figure}

A central contribution of BemaGANv2 is the introduction of the Multi-Envelope Discriminator (MED)\cite{b5, b6}, a novel discriminator architecture proposed by our research group. Unlike traditional discriminators that operate directly on raw waveforms or spectrograms, the MED analyzes time-domain envelopes derived from the audio signal. This distinct approach allows it to focus on temporal energy patterns crucial for perceptual quality, such as prosodic variation.

\begin{algorithm}
\caption{Multi-Envelope Discriminator (MED) Forward Pass}
\begin{algorithmic}[1]
\STATE \textbf{Input:} Real audio $y$, Generated audio $\hat{y}$
\STATE \textbf{Output:} Discriminator outputs and feature maps

\STATE \textbf{Initialize:} Envelope extraction modes $M = \{-1, 0, 1, 300, 500\}$
\STATE \hspace{1cm} $-1$: Lower envelope, $0$: Original signal, $1$: Upper envelope
\STATE \hspace{1cm} $300, 500$: Butterworth low-pass filtered envelopes (Hz)
\STATE $D_{outputs}^{real} \gets []$, $D_{outputs}^{gen} \gets []$
\STATE $F_{maps}^{real} \gets []$, $F_{maps}^{gen} \gets []$

\FOR{each $m \in M$}
    \STATE $D_m \gets \text{DiscriminatorE}(mode=m)$
    \STATE $out_r, fmap_r \gets D_m(y)$
    \STATE $out_g, fmap_g \gets D_m(\hat{y})$
    \STATE Append $out_r$ to $D_{outputs}^{real}$
    \STATE Append $out_g$ to $D_{outputs}^{gen}$
    \STATE Append $fmap_r$ to $F_{maps}^{real}$
    \STATE Append $fmap_g$ to $F_{maps}^{gen}$
\ENDFOR

\State \Return $D_{outputs}^{real}$, $D_{outputs}^{gen}$, $F_{maps}^{real}$, $F_{maps}^{gen}$
\end{algorithmic}
\end{algorithm}

These envelopes are extracted using the \texttt{Envelope()} function with configurations $F = \{-1, 0, 1, 300, 500\}$.
For $f_{\max} \in \{-1, 1\}$, Hilbert transform-based lower and upper envelopes are computed as $-|\mathcal{H}(-x(t))|$ and $|\mathcal{H}(x(t))|$, respectively, providing strict amplitude bounds.
For $f_{\max} = 0$, the original signal is preserved as a baseline.
For $f_{\max} \in \{300, 500\}$ Hz, Butterworth low-pass filtering precedes envelope extraction, capturing smoothed amplitude variations at syllabic (300 Hz) and phonemic (500 Hz) temporal scales.
This multi-scale approach enables the discriminator to evaluate both mathematically rigorous envelope accuracy and perceptually relevant temporal patterns. The use of multiple cutoff frequencies enables a multi-scale analysis of the audio's temporal structure, providing richer information about phrasing and amplitude modulation. As shown in Figure~\ref{fig2}, each extracted envelope is then processed by a dedicated 1D convolutional stack. The outputs from these stacks are subsequently aggregated to form the discriminator's final output. This design enables the model to detect patterns related to phrasing, prosody, and amplitude modulation properties, which are closely linked to human perception of naturalness. Consequently, the MED plays a crucial role in enhancing long-term temporal coherence in the generated audio

\begin{algorithm}
\caption{Envelope Extraction Methods}
\begin{algorithmic}[1]
\STATE \textbf{Input:} Audio signal $x$, Envelope mode $m$
\STATE \textbf{Output:} Extracted envelope signal

\IF{$m = -1$}
    \STATE $envelope \gets -|\text{Hilbert}(x)|$ \COMMENT{Lower envelope}
\ELSIF{$m = 0$}
    \STATE $envelope \gets x$ \COMMENT{Original signal}
\ELSIF{$m = 1$}
    \STATE $envelope \gets |\text{Hilbert}(x)|$ \COMMENT{Upper envelope (instantaneous amplitude)}
\ELSIF{$m = 300$ or $m = 500$}
    \STATE $x_{filtered} \gets \text{Butterworth\_LPF}(x, cutoff=m)$ \COMMENT{Low-pass filter}
    \STATE $envelope \gets |\text{Hilbert}(x_{filtered})|$ \COMMENT{Filtered envelope}
\ENDIF

\STATE \Return $envelope$
\end{algorithmic}
\end{algorithm}

The five envelope extraction modes serve distinct purposes in capturing temporal dynamics:

\begin{itemize}
    \item \textbf{Lower Envelope ($m=-1$):} Computed as the negative of the absolute Hilbert transform, tracking minimum amplitude variations and capturing the lower bound of signal energy.
    
    \item \textbf{Original Signal ($m=0$):} The unfiltered input signal serves as a baseline, providing reference discrimination without envelope-based preprocessing.
    
    \item \textbf{Upper Envelope ($m=1$):} The instantaneous amplitude obtained via absolute Hilbert transform, tracking maximum amplitude variations and representing the upper bound of signal energy.
    
    \item \textbf{300 Hz Filtered Envelope ($m=300$):} A Butterworth low-pass filter at 300 Hz isolates low-frequency components, with the envelope extracted via Hilbert transform to capture slow prosodic variations.
    
    \item \textbf{500 Hz Filtered Envelope ($m=500$):} A Butterworth low-pass filter at 500 Hz captures mid-low frequency dynamics, enabling discrimination of intermediate temporal patterns between phoneme-level and syllable-level structures.
\end{itemize}

This multi-scale envelope analysis enables the MED to detect subtle temporal characteristics across different frequency ranges, enhancing its sensitivity to prosodic variation, phrasing, and amplitude modulation—all of which are crucial for perceptual naturalness in long-term audio generation.

\subsection{Multi-Resolution Discriminator (MRD)}

To complement Multi-Envelope Discriminator (MED), BemaGANv2 also incorporates a Multi-Resolution Discriminator (MRD). This module operates in the time-frequency domain using STFT-based log-magnitude spectrograms. By processing multiple spectrograms computed with different FFT sizes and hop lengths, Multi-Resolution Discriminator (MRD) enforces spectral consistency across various resolutions. This multi-resolution analysis allows the discriminator to effectively capture both low-frequency pitch accuracy and high-frequency timbral details.

The Multi-Resolution Discriminator (MRD) primarily helps the model maintain harmonic sharpness and timbral accuracy, which are important for high-fidelity generation. Together with the MED, the MRD provides complementary perspectives—temporal envelope and spectral structure, respectively—thereby guiding the generator to produce perceptually convincing waveforms that are both rhythmically natural and spectrally accurate.

\subsection{Overall Architecture}

\begin{figure}[htbp]
\centering
\includegraphics[width=\columnwidth]{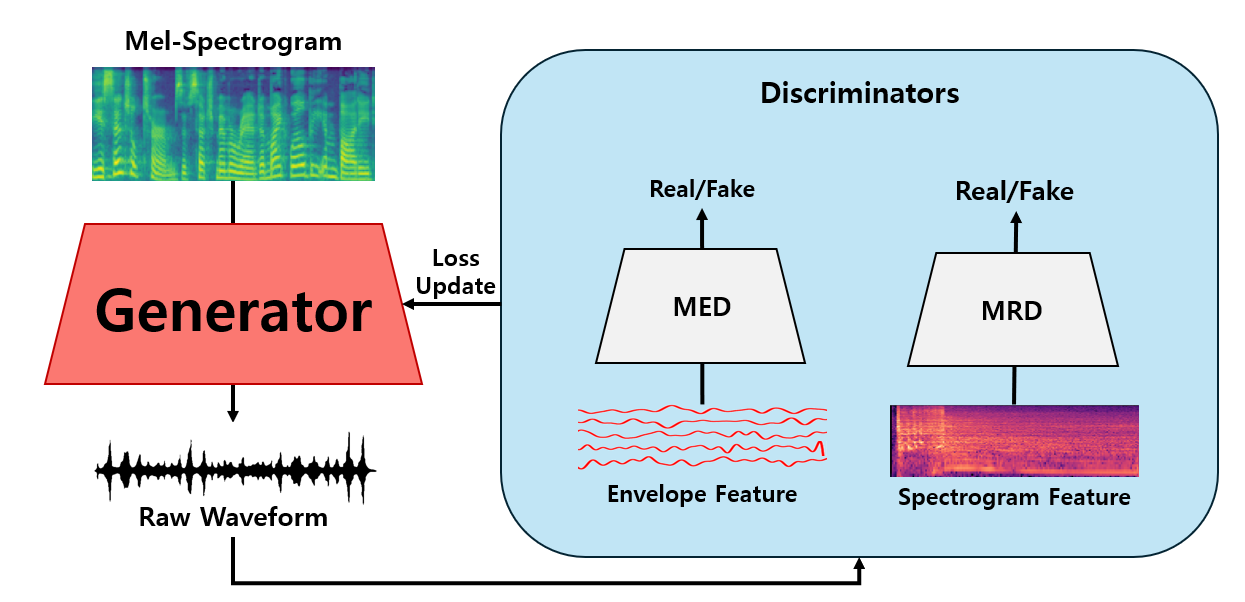}
\caption{Overview of the BemaGANv2 architecture. The generator converts Mel-spectrograms into raw waveforms, which are evaluated by two discriminators: the Multi-Envelope Discriminator (MED) and the Multi-Resolution Discriminator (MRD). Adversarial and auxiliary losses are backpropagated from both discriminators to train the generator.}
\label{fig3}
\end{figure}

Figure~\ref{fig3} presents the complete BemaGANv2 architecture. The generator transforms the input Mel-spectrogram into a raw waveform using AMP blocks. The output waveform is then assessed by both Multi-Envelope Discriminator (MED) and Multi-Resolution Discriminator (MRD). The Multi-Envelope Discriminator (MED) provides feedback based on temporal energy flow, focusing on characteristics like phrasing and prosody, while the Multi-Resolution Discriminator (MRD) provides feedback based on spectral consistency across various resolutions, ensuring harmonic sharpness and timbral accuracy. This dual-discriminator setup enables BemaGANv2 to generate audio that is both rhythmically natural and spectrally accurate. Adversarial and auxiliary losses are backpropagated from both discriminators to train the generator.

\subsection{Training Loss}

The training loss in BemaGANv2 follows the strategy proposed in HiFi-GAN and is based on the LSGAN formulation. Unlike the original GAN objective, which uses binary cross-entropy, LSGAN replaces it with a least squares loss to improve training stability and maintain non-vanishing gradients\cite{b14}.

\textbf{Adversarial Loss:} Let $D$ denote the discriminator and $G$ the generator. The discriminator is trained to assign real samples a value close to 1 and generated samples a value close to 0. The generator is optimized so that the generated samples fool the discriminator, i.e., are classified close to 1.
\begin{equation}
\mathcal{L}_{Adv}(D; G) = \mathbb{E}_{x,s}[(D(x) - 1)^2 + (D(G(s)))^2]
\end{equation}

where:
\begin{itemize}
    \item $x$ denotes real audio waveforms sampled from the training distribution;
    \item $s$ denotes input conditioning (Mel-spectrograms);
    \item $G(s)$ represents generated audio waveforms;
    \item $D(\cdot)$ represents the discriminator output;
    \item $\mathbb{E}_{x,s}[\cdot]$ denotes expectation over the joint distribution of real audio and conditioning.
\end{itemize}

\begin{equation}
\mathcal{L}_{Adv}(G; D) = \mathbb{E}_{s}[(D(G(s)) - 1)^2]
\end{equation}

\textbf{Mel-Spectrogram Loss:} We adopt an $L_1$ loss between the Mel-spectrograms of the real and generated waveforms:
\begin{equation}
\mathcal{L}_{Mel}(G) = \mathbb{E}_{x,s}[\| \phi(x) - \phi(G(s)) \|_1]
\end{equation}
where $\phi(\cdot)$ is the Mel-spectrogram transformation.

\textbf{Feature Matching Loss:} This loss minimizes the $L_1$ distance between intermediate features of the discriminator for real and generated samples:
\begin{equation}
\mathcal{L}_{FM}(G; D) = \sum_{l=1}^L \mathbb{E}[\| D^{(l)}(x) - D^{(l)}(G(s)) \|_1]
\end{equation}

\textbf{Total Losses:} The final loss functions for the generator and discriminator are given as:
{\small
\begin{equation}
\mathcal{L}_G = \sum_{k=1}^{K} [ \mathcal{L}_{Adv}(G; D_k) + \lambda_{fm} \mathcal{L}_{FM}(G; D_k) ] 
+ \lambda_{mel} \mathcal{L}_{Mel}(G)
\end{equation}
}
\begin{equation}
\mathcal{L}_D = \sum_{k=1}^{K} \mathcal{L}_{Adv}(D_k; G)
\end{equation}
where $K$ is the number of sub-discriminators, and $\lambda_{fm}, \lambda_{mel}$ are hyperparameters for feature matching and mel-spectrogram losses, respectively. These weights balance the contributions of the adversarial, feature matching, and mel-spectrogram losses, guiding the generator to synthesize high-quality and perceptually accurate audio.

In our implementation, we follow the HiFi-GAN protocol with the following loss weights:
\begin{itemize}
    \item $\lambda_{\mathrm{fm}} = 2$ \quad (feature matching loss weight)
    \item $\lambda_{\mathrm{mel}} = 45$ \quad (mel-spectrogram reconstruction loss weight)
\end{itemize}
These values are identical to those used in HiFi-GAN~\cite{b2} and have been validated across multiple GAN-based vocoder architectures. Complete training hyperparameters are provided in ~\ref{D}.2.

Our loss formulation is identical to HiFi-GAN\cite{b2}, which adopts the Least-Squares GAN (LSGAN) objective\cite{b14} for improved training stability. The key difference lies in the discriminator architecture (MED+MRD in BemaGANv2 vs. MPD+MSD in HiFi-GAN), while the loss weighting scheme remains unchanged

\section{Experiments and Results} \label{sec4}

In this section, we conduct a comprehensive evaluation of various discriminator configurations and their impact on audio reconstruction quality. We place particular emphasis on the architectural combinations used in BemaGANv2, while also referencing models such as HiFi-GAN, BigVGAN, and our previous BemaGAN. The goal is not only to compare model performance, but also to offer practical insights for vocoder design choices in short-term and long-term audio synthesis tasks. This section is structured as follows: we first outline the evaluation metrics used, then present the objective and subjective results, followed by a qualitative comparison of Mel-spectrogram structures, and finally summarize our key findings. For short-term evaluation, we used audio clips of approximately 20 seconds, including a diverse range of content such as human speech, drum and percussion sounds, short musical phrases, and car engine noise. For long-term evaluation, we used 90-second segments of full music tracks spanning various genres. It is important to note that both the short-term and long-term test sets were intentionally constructed to evaluate out-of-distribution (OOD) generalization, as all test audio—including environmental sounds, musical instruments, and full music tracks—differs substantially from the single-speaker speech in the LJSpeech training corpus. This represents a significant domain shift and provides a rigorous test of each model's extrapolation capabilities beyond the training distribution.
To ensure a fair comparison of discriminator configurations, all models
except the original HiFi-GAN employed a generator architecture based on
AMP blocks with embedded Snake activation functions. All models in the
comparative study were trained for 500 epochs under identical conditions
(detailed in Appendix ~\ref{D}.7). Following the established findings from
HiFi-GAN~\cite{b2} and BigVGAN~\cite{b3}, which
demonstrated that multi-discriminator configurations significantly
outperform single-discriminator setups, our primary experiments focus on
evaluating novel combinations of discriminators (MED+MRD, MSD+MED, etc.).

In addition, to address the need for empirical validation of individual
component contributions, we conducted two targeted ablation experiments.
First, we evaluated a \textbf{MED-only} configuration, where the
AMP+Snake generator is paired with MED as the sole discriminator. This
isolates MED's independent contribution to audio quality, particularly
in temporal envelope modeling. Second, we applied the HiFi-GAN
discriminator combination (\textbf{MPD+MSD}) to the same AMP+Snake
generator used across all other models. This configuration, denoted as
\textbf{HiFi-GAN w/ AMP+Snake} in our tables, strengthens variable
control by holding the generator constant and allowing readers to
directly compare the effect of different discriminator configurations.
Note that a detailed ablation study on the Snake activation function was
previously conducted by Lee et al.~\cite{b3}, where Snake
was shown to significantly improve periodicity modeling compared to
Leaky ReLU. Rather than duplicating this established result, the
inclusion of the HiFi-GAN w/ AMP+Snake experiment indirectly
illustrates the generator-side contribution by contrasting it with the
original HiFi-GAN, which uses the Leaky ReLU-based ResBlock generator.

It is important to emphasize that the goal of our comparative
experiments is not to identify which single discriminator performs best
in isolation, but rather to demonstrate that \textit{how discriminators
are combined} matters as much as---if not more than---the choice of
individual discriminators. The ablation experiments complement this
primary objective by providing empirical evidence for the individual
roles of key components.

\subsection{Evaluation Metrics}
To ensure both analytical rigor and perceptual validity, we adopt four objective and two subjective metrics:

\textbf{Objective Metrics}
\begin{itemize}
    \item \textbf{Fréchet Audio Distance (FAD)}: Quantifies the distributional divergence between the generated and the ground-truth audio embeddings\cite{b15}.
    \item \textbf{Structural Similarity Index (SSIM)}: Evaluates the perceptual closeness between Mel-spectrogram images across RGB channels\cite{b16}.
    \item \textbf{Pearson Correlation Coefficient (PCC)}: Measures the degree of linear correlation in frequency domain characteristics\cite{b17}.
    \item \textbf{Mel-Cepstral Distortion (MCD)}: Captures the distortion in MFCC sequences using Dynamic Time Warping\cite{b18}.
    \item \textbf{Multi-Resolution STFT Loss (M-STFT):} Measures spectral
    reconstruction quality across multiple STFT resolutions~\cite{b42}.
    \item \textbf{Periodicity:} Evaluates the preservation of periodic
    structures in the generated waveform by measuring pitch periodicity
    estimation error~\cite{b43}.
\end{itemize}

Among these metrics, M-STFT and Periodicity are adopted following the evaluation protocol established by BigVGAN~\cite{b3}, providing a standardized framework for spectral reconstruction quality and harmonic structure preservation that complements the existing perceptual and distributional metrics.

We note that certain metrics used in BigVGAN---specifically PESQ (Perceptual Evaluation of Speech Quality, ITU-T P.862) and V/UV F1 (Voiced/Unvoiced F1 score)---were not included in our evaluation. Both metrics are designed exclusively for speech signals: PESQ models the human auditory system's sensitivity to speech distortions, while V/UV F1 measures classification accuracy of voiced and unvoiced segments. Since our test sets include musical content, environmental sounds, and other non-speech audio, these speech-specific metrics would not yield reliable or interpretable results across the entire evaluation set.

Regarding FAD, while it remains widely used for evaluating generative audio models, recent studies have identified several limitations. Gui et al.~\cite{b41} demonstrated that FAD scores are susceptible to sample size bias, sensitive to the choice of audio embedding model, and influenced by the quality and composition of the reference set. Their findings further showed that commonly used configurations---such as VGGish embeddings with the MusicCaps reference set---may not correlate well with subjective quality judgments. In light of these concerns, we retain FAD as a supplementary distributional measure for completeness and comparability with prior work, but avoid overreliance on FAD as primary evidence of perceptual quality.

In our implementation, we adopt the standard SPTK-based configuration and compute MCD using fastDTW after extracting the mel-cepstrum from each frame. It is important to note that while typical MCD values for high-quality speech synthesis are often below 1.0, the reported MCD values in our tables, may exceed 1. This is attributed to practical constraints, such as limited computing resources, short training epochs because of the need to evaluate across six different model configurations. However, we empirically confirmed that when tested under typical conditions used in prior work—namely short-form utterances of around 10 seconds and with sufficiently trained models—our implementation yields MCD values well below 1.0. This validates the effectiveness and correctness of our MCD implementation for comparative analysis across the evaluated models.

SSIM, originally designed for image comparison, has been successfully adapted for audio evaluation by measuring structural similarity between mel-spectrogram representations \cite{b37, b38}. It effectively captures spectral-temporal coherence and has been validated in recent vocoder benchmarks.

Similarly, we use Pearson Correlation Coefficient (PCC) which measures correlation between ground-truth and reconstructed mel-spectrograms. PCC has been widely adopted in audio reconstruction tasks \cite{b39, b40} and correlates well with perceptual similarity.

\textbf{Subjective Metrics}
\begin{itemize}
    \item \textbf{MOS (Mean Opinion Score)}: Participants rate the overall perceptual quality of the audio samples without seeing the ground truth.
    \item \textbf{SMOS (Similarity MOS)}: Participants compare the similarity between the generated samples and the reference ground-truth audio.
\end{itemize}

All subjective evaluations are conducted on a 5-point scale with 95\% confidence intervals, using samples from both short- and long-term audio.

\subsection{Objective Evaluation Results}
\begin{table*}[htbp]
\centering
\caption{Short-Term Audio Objective Metrics}
\begin{adjustbox}{max width=\textwidth}
\begin{tabular}{l|c|c|c|c|c|c}
\hline
\textbf{Models} & \textbf{FAD $\downarrow$} & \textbf{SSIM $\uparrow$} & \textbf{PCC $\sim 1$} & \textbf{MCD $\downarrow$} & \textbf{M-STFT $\downarrow$} & \textbf{Periodicity $\downarrow$} \\ \hline
\textbf{BemaGANv2 (MED + MRD)}       & \textbf{0.911} & \textbf{0.85} & \textbf{0.949} & \textbf{1.652} & \textbf{1.262} & 0.161 \\
\quad MED only                        & 1.116  & 0.82 & 0.94  & 1.887 & 1.453 & 0.2208 \\ \hline
BemaGAN (MPD + MED)                   & 3.194  & 0.68 & 0.861 & 3.18  & 1.893 & 0.1883 \\ \hline
BigVGAN (MPD + MRD)                   & 1.906  & 0.79 & 0.906 & 2.56  & 1.519 & 0.1716 \\ \hline
HiFi-GAN (MPD + MSD)                  & 6.06   & 0.73 & 0.792 & 3.22  & 3.256 & 0.1883 \\
\quad w/ AMP + Snake                  & 1.5363 & 0.78 & 0.888 & 2.553 & 1.527 & \textbf{0.1543} \\ \hline
MSD + MED                             & 2.337  & 0.67 & 0.879 & 2.964 & 1.803 & 0.1735 \\ \hline
MSD + MRD                             & 3.327  & 0.63 & 0.841 & 3.354 & 1.948 & 0.1802 \\ \hline
MED + MPD + MRD                       & 2.146  & 0.71 & 0.851 & 3.136 & 1.771 & 0.1692 \\ \hline
\end{tabular}
\end{adjustbox}
\label{tab:short_objective}
\end{table*}

\begin{table*}[htbp]
\centering
\caption{Long-Term Audio Objective Metrics}
\begin{adjustbox}{max width=\textwidth}
\begin{tabular}{l|c|c|c|c|c|c}
\hline
\textbf{Models} & \textbf{FAD $\downarrow$} & \textbf{SSIM $\uparrow$} & \textbf{PCC $\sim 1$} & \textbf{MCD $\downarrow$} & \textbf{M-STFT $\downarrow$} & \textbf{Periodicity $\downarrow$} \\ \hline
\textbf{BemaGANv2 (MED + MRD)}       & 2.681 & \textbf{0.78} & \textbf{0.945} & \textbf{1.8} & \textbf{1.5141} & \textbf{0.1235} \\
\quad MED only                        & \textbf{2.204}  & 0.75 & 0.945 & 1.966 & 1.638 & 0.1361 \\ \hline
BemaGAN (MPD + MED)                   & 6.612  & 0.63 & 0.866 & 2.841 & 1.827 & 0.1657 \\ \hline
BigVGAN (MPD + MRD)                   & 3.58   & 0.71 & 0.908 & 2.28  & 1.613 & 0.1504 \\ \hline
HiFi-GAN (MPD + MSD)                  & 30.883 & 0.08 & 0.702 & 7.75  & 3.115 & 0.2178 \\
\quad w/ AMP + Snake                  & 4.274  & 0.69 & 0.885 & 2.392 & 1.622 & 0.1483 \\ \hline
MSD + MED                             & 6.66   & 0.63 & 0.810 & 2.92  & 1.7301 & 0.1922 \\ \hline
MSD + MRD                             & 7.221  & 0.60 & 0.776 & 3.345  & 1.75 & 0.1793 \\ \hline
MED + MPD + MRD                       & 5.969  & 0.66 & 0.878 & 2.7557 & 1.6801 & 0.1762 \\ \hline
\end{tabular}
\end{adjustbox}
\label{tab:long_objective}
\end{table*}

As presented in Table~\ref{tab:short_objective} and Table~\ref{tab:long_objective}, BemaGANv2 (MED+MRD) consistently achieves superior performance across most objective metrics for both short- and long-term audio generation. BemaGANv2 outperforms all baselines in FAD, SSIM, PCC, MCD, and M-STFT for short-term audio, and achieves the best scores in SSIM, PCC, MCD, M-STFT, and Periodicity for long-term audio.

HiFi-GAN with its original generator (Leaky ReLU + ResBlock) shows significantly degraded performance in long-term audio across all metrics, which aligns with the anomalous waveform length doubling observed in our extended analysis (Section~\ref{sec5}). In contrast, when the same HiFi-GAN discriminator combination (MPD+MSD) is applied to the AMP+Snake generator (\textbf{HiFi-GAN w/ AMP+Snake}), performance improves substantially---e.g., long-term FAD drops from 30.883 to 4.274, and MCD from 7.75 to 2.392. This result provides strong evidence that the generator architecture, particularly the Snake activation and anti-aliasing mechanism, plays a critical role in long-term audio quality, consistent with the findings of Lee et al.~\cite{b3}. Notably, the duration doubling anomaly previously observed in HiFi-GAN was not present in this configuration, further supporting the generator-side origin of the issue.

\textbf{Ablation: MED-only configuration.} To isolate the contribution of MED, we evaluated a configuration using MED as the sole discriminator with the same AMP+Snake generator. The MED-only model achieves competitive results: in short-term evaluation, it attains the second-lowest M-STFT (1.453) and strong PCC (0.94), and in long-term evaluation, it achieves the lowest FAD (2.204) among all models. These results confirm that MED independently contributes meaningful supervisory signals for audio quality, particularly in distributional fidelity and temporal envelope modeling. However, compared to the full BemaGANv2 (MED+MRD), the MED-only configuration shows lower performance in SSIM (0.82 vs.\ 0.85 for short-term; 0.75 vs.\ 0.78 for long-term), M-STFT (1.453 vs.\ 1.126 for short-term), and Periodicity (0.2208 vs.\ 0.161 for short-term; 0.1361 vs.\ 0.1235 for long-term). This demonstrates that MRD provides complementary spectral supervision that enhances structural consistency and harmonic preservation---capabilities that MED's temporal envelope analysis alone does not fully cover. The synergy between MED and MRD is thus empirically validated: MED excels at temporal envelope and distributional fidelity, while MRD strengthens spectral sharpness accuracy.

While our earlier experiments indicated that HiFi-GAN performed better in short-term tasks\cite{b6}, additional training with extended epochs has revealed a different trend. BemaGANv2, when trained for a longer duration, not only maintains its superior performance in long-term audio generation but also clearly outperforms other models in short-term tasks. This trend may be attributed to the gradient properties of the Snake activation function. Unlike Leaky ReLU, which maintains a non-zero gradient across its domain, Snake contains regions where the derivative approaches zero, potentially slowing convergence in early training stages. However, with sufficient training, the model leverages the periodic inductive bias of Snake to capture harmonic and transient features more effectively. A detailed mathematical explanation of this behavior, including derivative analysis and its implications for convergence, is provided in ~\ref{A}. 

To verify the robustness of these findings, we additionally trained BemaGANv2 with four different random seeds and report the mean and standard deviation of all objective metrics. The results, presented in Appendix~\ref{E}, confirm that BemaGANv2 exhibits low variability across seeds, with most metrics showing small standard deviations, indicating that the reported performance advantages are reproducible and not dependent on a particular random initialization.

\subsection{Subjective Evaluation Results}

Table~\ref{tab:mos} and Table~\ref{tab:smos} report the results of human evaluations, which are broadly consistent with the objective measurements. These results provide further evidence that architectural decisions—such as the choice of activation functions and discriminator types—impact the perceptual quality across different audio durations. The MOS and SMOS evaluations were conducted with 137 general participants recruited via crowdsourcing. Each audio sample was rated by at least 15 listeners. All reported scores include 95\% confidence intervals(CI) as shown in Table~\ref{tab:mos} and Table~\ref{tab:smos}. We used general listeners rather than expert listeners to ensure our evaluation reflects real-world user perception, which is the ultimate goal of high-fidelity audio generation systems.

\begin{table}[htbp]
\centering
\caption{Mean Opinion Score (MOS)}
\resizebox{\linewidth}{!}{%
\begin{tabular}{l|c|c}
\hline
\textbf{Models} & \textbf{Short Audio MOS $\uparrow$} & \textbf{Long Audio MOS $\uparrow$} \\ \hline
Ground Truth           & 3.67(±0.09) & 4.73(±0.12) \\ \hline
BemaGANv2           & \textbf{3.08(±0.10)} & \textbf{3.46}(±0.12) \\ \hline
BemaGAN             & 1.74(±0.11) & 2.67(±0.13) \\ \hline
BigVGAN             & 2.52(±0.14) & 3.07(±0.13) \\ \hline
HiFi-GAN            & 2.91(±0.12) & 1.14(±0.09) \\ \hline
MSD + MED           & 2.16(±0.09) & 2.313(±0.14) \\ \hline
MSD + MRD           & 2.14(±0.19) & 2.37(±0.15) \\ \hline
MED + MPD + MRD     & 2.17(±0.09) & 2.31(±0.12) \\ \hline
\end{tabular}%
}
\label{tab:mos}
\end{table} 

\begin{table}[htbp]
\centering
\caption{Similarity Mean Opinion Score (SMOS)}
\resizebox{\linewidth}{!}{%
\begin{tabular}{l|c|c}
\hline
\textbf{Models} & \textbf{Short Audio SMOS $\uparrow$} & \textbf{Long Audio SMOS $\uparrow$} \\ \hline
BemaGANv2           & \textbf{3.07(±0.09)} & \textbf{3.53}(±0.09) \\ \hline
BigVGAN             & 2.61(±0.12) & 3.09(±0.09) \\ \hline
HiFi-GAN            & 2.66(±0.12)  & 1.03(±0.03)  \\ \hline
\end{tabular}%
}
\label{tab:smos}
\end{table}

Consistent with the objective evaluation results, BemaGANv2 received the highest subjective scores in both perceptual quality (MOS) and similarity (SMOS), further validating its effectiveness across different audio durations.

\subsection{Comparing Mel-spectrogram Structure}

\begin{figure*}[t]
\centering
\includegraphics[width=\textwidth]{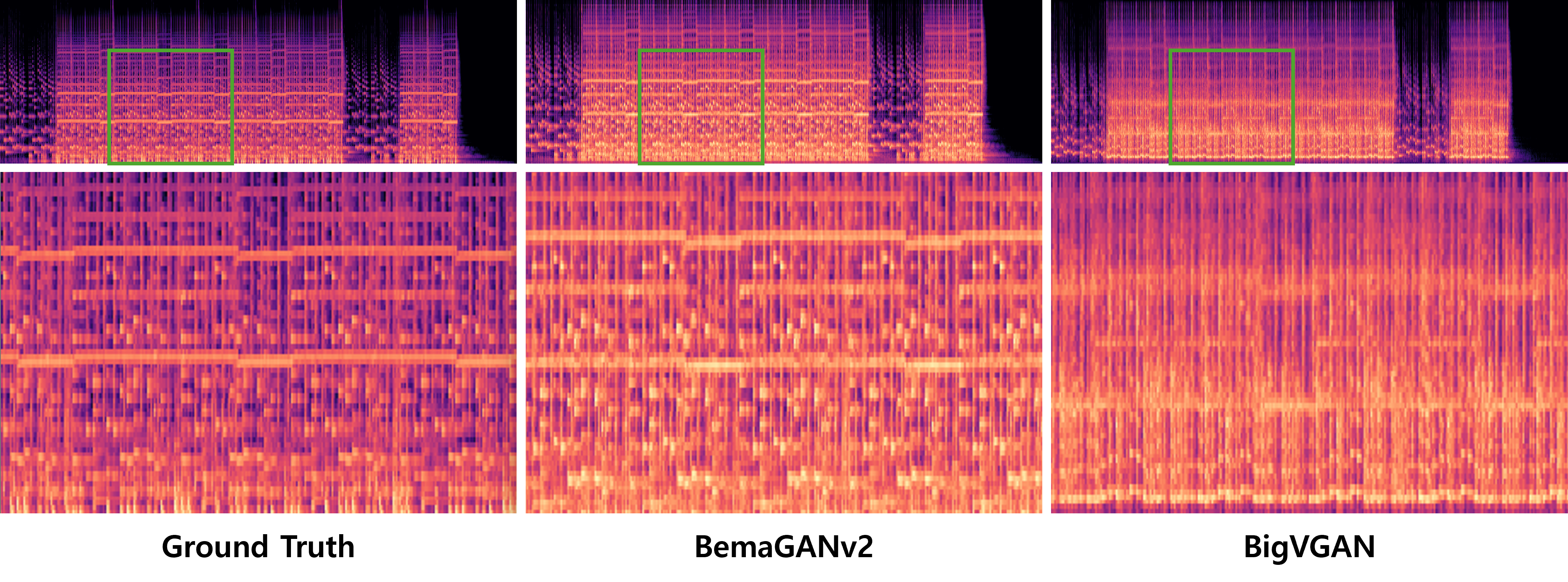}
\caption{Mel-Spectrogram visualization of samples from Ground Truth, BigVGAN, and BemaGANv2 trained on LJSpeech.}
\label{fig:melvis}
\end{figure*}

The visualization of the Mel spectrogram in Figure~\ref{fig:melvis} reinforces these findings. BemaGANv2 shows better structure preservation and noise suppression in long-term audio compared to BigVGAN. 

\subsection{Result}

The results clearly demonstrate that BemaGANv2 achieves superior performance in both short- and long-term audio tasks, outperforming all other models across objective and subjective evaluations. This highlights the effectiveness of its generator design and the MED+MRD discriminator configuration, particularly in preserving both high-frequency details and long-term periodic structures. The addition of ablation configurations—MED only and HiFi-GAN with the AMP+Snake generator—further confirms that the MED+MRD combination provides the most balanced supervisory signal, and that the generator architecture plays a critical role in long-term generation stability.

Interestingly, although the MED+MPD+MRD combination yielded competitive objective metrics—especially in long-term evaluation—it showed significantly lower subjective scores. Through an analysis of training and validation loss curves (Appendix B), we confirmed that this is not due to overfitting. Instead, we attribute the degradation in perceptual quality to mode collapse, likely caused by adversarial instability when employing an excessive number of discriminators. This observation, combined with the ablation results showing that even the MED-only configuration achieves reasonable performance, suggests that discriminator design should prioritize complementary coverage of perceptual dimensions over sheer quantity. These findings highlight a crucial practical insight: the importance of carefully balancing feedback diversity and adversarial stability in discriminator design, cautioning against simply adding more discriminators without considering their synergistic effects.

For further experimental details, including loss curves and configuration analysis, as well as complete implementation specifications—such as hardware settings, all hyperparameters, preprocessing pipelines, and training configurations—please refer to ~\ref{B} and ~\ref{D}. All code, configuration files, and pretrained models are publicly available at the link mentioned in the abstract.

\section{Discussion} \label{sec5}

During the evaluation of vocoder outputs, we observed a peculiar issue in the HiFi-GAN model where long-term audio samples resulted in outputs nearly double the expected length. This phenomenon was not observed in other models (e.g., BigVGAN, BemaGANv2), despite identical training data, architectures, and preprocessing pipelines.

To identify the source of this anomaly, we systematically investigated several components of the inference pipeline:
\begin{itemize}
    \item Configuration consistency between training and inference (e.g., \texttt{hop\_size}, \texttt{n\_fft}) — confirmed to be consistent.
    \item STFT and Mel-spectrogram center alignment — identical across models and not responsible for the anomaly.
    \item Dataset bias — all models were trained on LJSpeech, which consists of short, speech-based audio samples.
    \item Model architecture — no differences except the generator architecture. 
\end{itemize}

Among the models evaluated in our comparative study, HiFi-GAN is the only one that uses the Leaky ReLU-based ResBlock generator, while all other models employ the AMP+Snake generator architecture. This observation initially suggested the activation function as a potential factor in the anomaly. To investigate this further, we conducted an additional experiment in which HiFi-GAN's discriminator combination (MPD+MSD) was applied to the same AMP+Snake generator used across all other models. Notably, when using this configuration (HiFi-GAN w/ AMP+Snake), the waveform duration doubling anomaly was not observed, and long-term audio quality improved substantially. This result provides strong evidence suggesting that the generator architecture---particularly the choice of activation function and the presence of anti-aliasing---plays a significant role in the observed anomaly, rather than the discriminator configuration.

These findings underscore the importance of periodic inductive bias in vocoders tasked with long-term synthesis. The Snake activation function, which provides learnable periodicity through its oscillatory formulation~\cite{b4}, appears to enable more stable extrapolation of periodic structures beyond the training distribution of short-length signals. However, we note that the AMP+Snake generator differs from HiFi-GAN's original generator in multiple aspects---including both the activation function and the anti-aliasing mechanism---and therefore, while these results substantially narrow the source of the anomaly to the generator side, a fully isolated single-variable experiment (e.g., changing only the activation function while keeping all other architectural components identical) would be needed to establish definitive causality. Nevertheless, in all of our experiments, Snake-based generators consistently exhibited better stability and fidelity in long-term audio generation.

Beyond the HiFi-GAN anomaly discussed above, we also observed performance limitations of BemaGANv2 in certain audio conditions. Specifically, BemaGANv2 achieves high reconstruction quality on relatively simple audio sources such as solo instruments and single vocal tracks, but exhibits noticeable quality degradation when processing complex polyphonic audio with multiple simultaneously playing instruments (e.g., full band arrangements, orchestral passages). We attribute this to the increased harmonic overlap and spectral density in such signals, which pose challenges for the envelope-based discrimination provided by MED, combined with the fact that the model was trained on LJSpeech, which consists exclusively of single-speaker speech. Addressing this limitation through training on more diverse and polyphonic datasets represents an important direction for future work.

For reproducibility and further inspection, debugging logs and comparative duration graphs are provided in the ~\ref{C}

\section{Conclusion}
In this study, we presented BemaGANv2, a vocoder designed to enhance both the fidelity and temporal coherence of waveform reconstruction. Motivated by theoretical considerations and validated through extensive empirical evaluation, BemaGANv2 was shown to outperform existing GAN-based vocoders such as HiFi-GAN and BigVGAN, particularly in long-term audio generation. Its generator, based on AMP blocks, and the proposed Multi-Envelope Discriminator (MED), effectively capture both local signal detail and long-range periodic structure.

Our results demonstrate that BemaGANv2 achieves superior performance across both objective and subjective evaluation metrics, in both short- and long-form audio scenarios.

\textbf{Key contributions of this work include:}
\begin{itemize}
    \item A tutorial-style analysis of GAN-based vocoder evolution, from MelGAN to BigVGAN.
    \item The design and integration of the Multi-Envelope Discriminator (MED) to enhance prosodic and envelope-level modeling.
    \item A comparative evaluation of six discriminator configurations, shedding light on architectural trade-offs.
    \item An empirical finding linking activation functions to inference stability in long-term generation tasks.
\end{itemize}

These findings suggest that BemaGANv2 is a promising vocoder for Text-to-Music (TTM) and Text-to-Audio (TTA) models, particularly in diffusion-based generation pipelines. As future work, we plan to integrate BemaGANv2 into multimodal generative AI systems to enable more expressive and temporally coherent audio synthesis. We anticipate that this direction will help bridge the gap between machine-generated and human-perceived audio realism.

A limitation of this study is that experiments were conducted primarily on the LJSpeech dataset, a single-speaker English corpus. While we attempted to scale to larger, multi-speaker datasets such as LibriTTS, the computational demands of our study—which requires training and evaluating multiple discriminator configurations under identical conditions—exceeded our available resources (a single NVIDIA A100 GPU via Google Colab, compared to the 8 NVIDIA V100 GPUs used in the BigVGAN study). Future work should include evaluation on multi-speaker and multi-language datasets such as LibriTTS and VCTK to further validate the generalizability of BemaGANv2 across diverse acoustic conditions.

\section*{Acknowledgments}
This work was supported by the National Research Foundation of Korea(NRF) grant funded by the Korea government(MSIT)(RS-2024-00352526)

\appendix
\renewcommand{\thefigure}{A.\arabic{figure}}
\renewcommand{\theequation}{A.\arabic{equation}}
\setcounter{figure}{0}
\setcounter{equation}{0}
\section{Analysis of the Activation Function}
\label{A}

Our earlier experiments suggested that HiFi-GAN performs best on short-term audio\cite{b6}, especially under limited training epochs. However, in the extended study presented in this paper, we provided sufficient training epochs, and the results indicate that BemaGANv2 outperforms all baselines in both short- and long-term waveform reconstruction tasks. In this section, we provide a mathematical interpretation to explain this behavior, particularly in relation to the activation functions used.

\subsection{Periodic Bias of the Snake Function}

Snake function is designed to introduce periodic inductive bias, which makes it highly effective at capturing repeating structures in long-term audio signals. However, in short audio segments—especially those dominated by non-periodic transient features or high-frequency details—this periodic bias may cause the model to underfit rapid variations, leading to degraded performance if not sufficiently trained.

\subsection{Gradient Vanishing Issue}

The derivative of the Snake function is:

\begin{equation}
f_\alpha'(x) = 1 + \sin(2\alpha x)
\end{equation}

When $\sin(2\alpha x) = -1$, this derivative periodically approaches zero. In such regions, the gradient flow is suppressed, making it difficult for the model to update its weights during training. This is especially problematic in learning small amplitude variations common in short-term signals, where precise gradient flow is crucial.

In contrast, Leaky ReLU\cite{b19} is defined as:

\begin{equation}
\mathrm{LeakyReLU}(x) =
\begin{cases}
x, & \text{if } x \geq 0 \\
\beta x, & \text{otherwise}
\end{cases}
\end{equation}

and its derivative is constant:

\begin{equation}
\mathrm{LeakyReLU}'(x) =
\begin{cases}
1, & \text{if } x \geq 0 \\
\beta, & \text{otherwise}
\end{cases}
\end{equation}

This consistent gradient ensures stable learning even in shallow networks or early training stages. Therefore, without enough training iterations, Snake-based generators may struggle to achieve the same level of short-term fidelity as those using Leaky ReLU. Our results suggest that when given sufficient epochs, the advantages of Snake’s periodic bias are fully realized, enabling BemaGANv2 to excel in both short- and long-term tasks.

\section{Loss Curve}
\renewcommand{\thefigure}{B.\arabic{figure}}
\renewcommand{\theequation}{B.\arabic{equation}}
\setcounter{figure}{0}
\setcounter{equation}{0}
\label{B}

\begin{figure}[htbp]
    \centering
    \includegraphics[width=\linewidth]{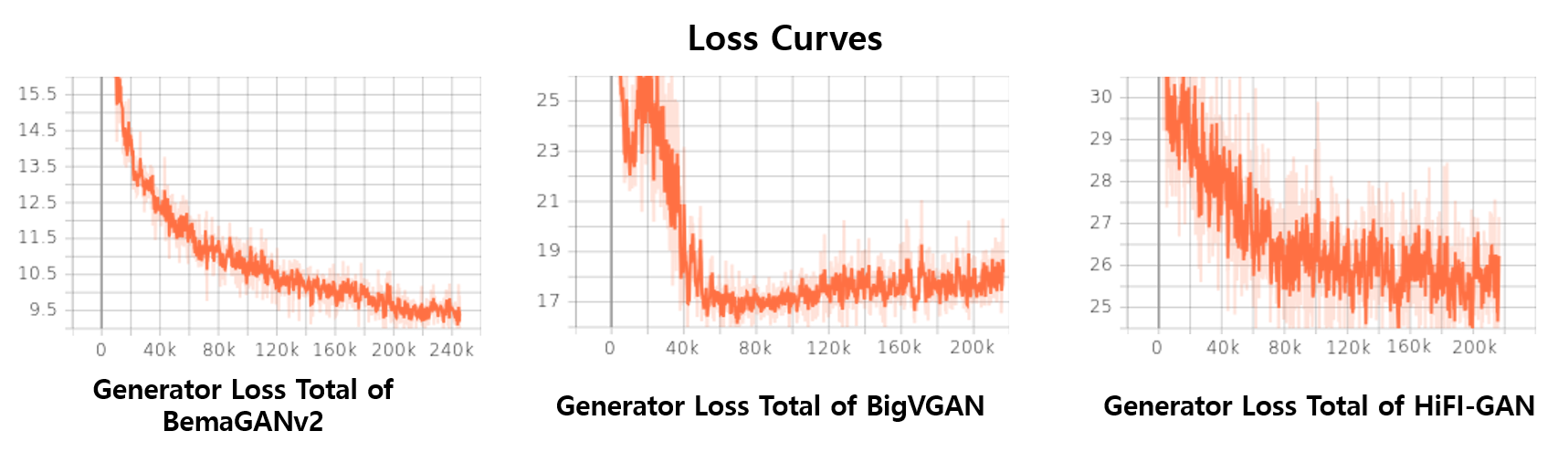}
    \caption{Training loss curves of BemaGANv2, HiFi-GAN, and BigVGAN.}
    \label{fig:loss_comparison}
\end{figure}

\begin{figure}[htbp]
    \centering
    \includegraphics[width=\linewidth]{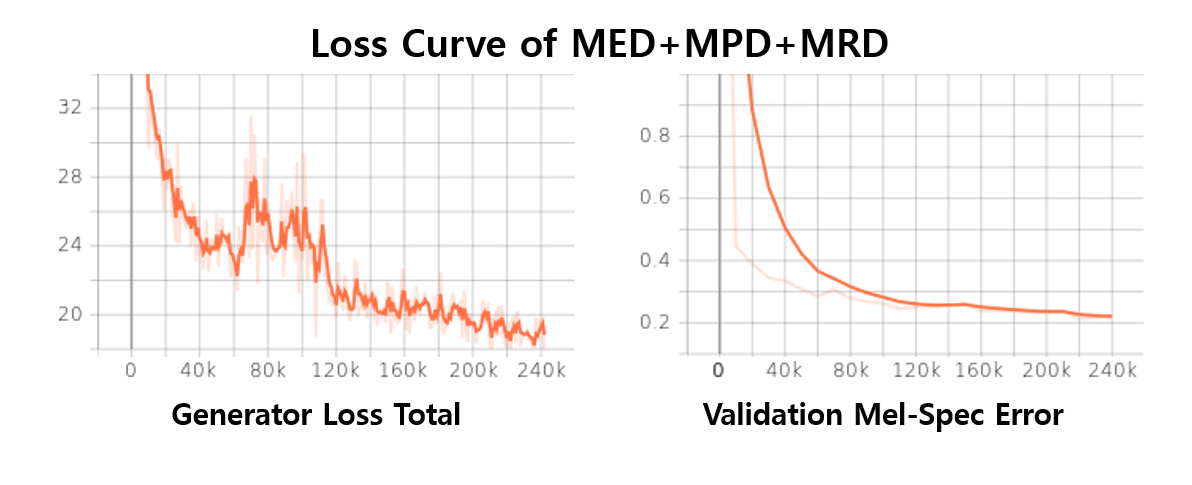}
    \caption{Training and validation loss curves of the MED+MPD+MRD combination.}
    \label{fig:loss_mmm}
\end{figure}

Figure~\ref{fig:loss_comparison} presents the training loss curves for BemaGANv2, HiFi-GAN, and BigVGAN. The results indicate that BemaGANv2 demonstrates superior training stability compared to the other models. This suggests that the combination of Multi-Envelope Discriminator (MED) and Multi-Resolution Discriminator (MRD) discriminators in BemaGANv2 provides more consistent adversarial feedback and contributes to robust convergence.

Figure~\ref{fig:loss_mmm} illustrates the training and validation loss curves for the MED+MPD+MRD configuration. Although previous studies raised concerns about potential overfitting in multi-discriminator setups, the curve clearly shows that the observed performance drop in subjective metrics is not due to overfitting. Rather, the consistency between training and validation curves supports our conclusion that the degradation is caused by mode collapse resulting from adversarial instability.

\section{Debugging Logs and Analysis}
\label{C}
\renewcommand{\thefigure}{C.\arabic{figure}}
\renewcommand{\theequation}{C.\arabic{equation}}
\setcounter{figure}{0}
\setcounter{equation}{0}

To diagnose the anomalous behavior observed during inference with HiFi-GAN—where long-term audio samples were generated at nearly double the intended length—we conducted a series of logging and validation procedures. This appendix provides sample logs and graphical summaries of the debugging process.

\subsection{Waveform Length Validation Log}

The following logs show the comparison between original and generated waveform lengths for multiple test samples. All lengths are computed in samples and seconds.

\begin{verbatim}
===== Inference: example_1.wav =====
Original sampling rate: 44100
Original waveform length: 4,279,739 samples
Original duration: 97.05 sec
Mel shape: torch.Size([1, 80, 16717])
Expected waveform length: 4,279,552 samples
Generated waveform length: 4,279,552 samples
Length difference: 0 samples (0.00 sec)

===== Inference: example_2.wav =====
Original waveform length: 3,873,082 samples
Generated waveform length: 3,873,024 samples
Difference: 0 samples (0.00 sec)
\end{verbatim}

\subsection{Length Consistency Visualization}

\begin{figure}[htbp]
    \centering
    \includegraphics[width=0.85\linewidth]{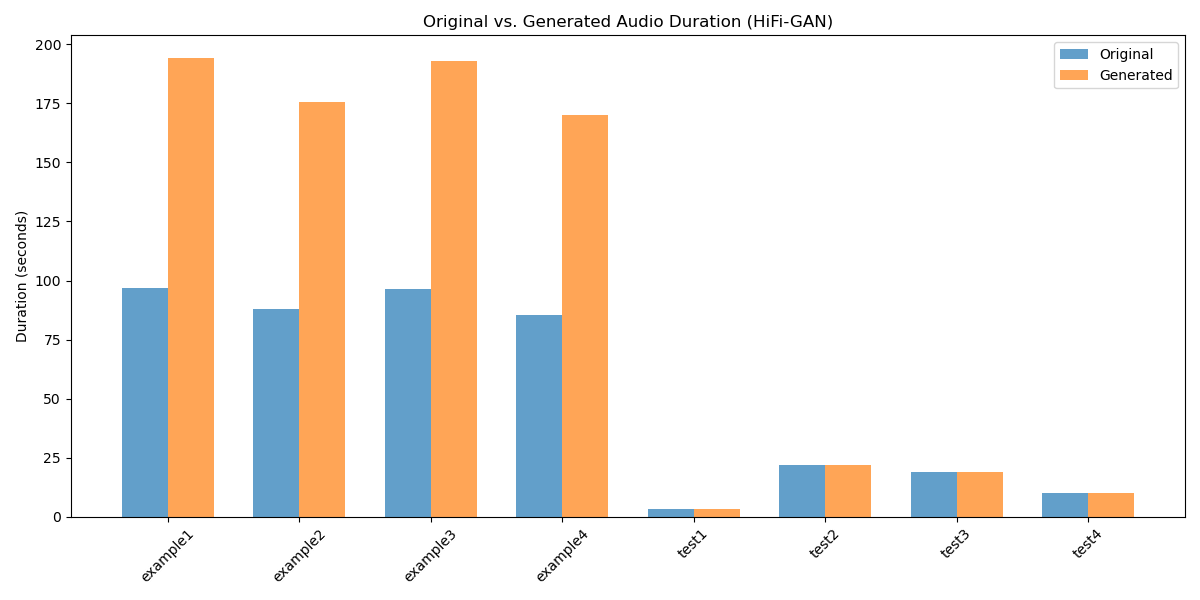}
    \caption{Comparison between original and generated waveform lengths for short and long audio inputs.}
    \label{fig:length_debug}
\end{figure}

Figure~\ref{fig:length_debug} confirms that waveform length discrepancies are not consistently present across all durations. Short clips (e.g., 20 seconds) are reconstructed with accurate length, while anomalies are observed predominantly in long-form audio in specific configurations.

\subsection{Mel-to-Audio Consistency Verification}

To verify the alignment between mel-spectrogram frame count and audio frame count, we tracked the following:

\begin{itemize}
    \item $\text{mel\_length} = \text{waveform\_length} / \text{hop\_size}$
    \item All inference samples conformed to this relationship.
\end{itemize}

\subsection{Conclusion}

After validating inference logs, comparing mel frame counts, and matching waveform durations, we ruled out issues in pre-processing, STFT padding, or inference time truncation. These observations suggest that the previously reported duration anomaly is not caused by implementation bugs, but likely arises from training data mismatch, activational bias (e.g., LeakyReLU vs. Snake), or temporal extrapolation failure specific to HiFi-GAN under long-term generation settings.

\section{Implementation Details}
\label{D}

This appendix provides detailed implementation specifications to ensure full reproducibility of our experiments. Note that architectural details and loss function formulations are described in Section 3 of the main paper.

\subsection{Computational Environment}

\begin{itemize}
    \item \textbf{Hardware:} NVIDIA A100 GPU via Google Colab
    \item \textbf{Software:} Python 3.9, PyTorch 1.13.0
    \item \textbf{Key Libraries:} librosa, scipy, numpy, tensorboard, pandas
\end{itemize}

\subsection{Training Hyperparameters}

All training hyperparameters are defined in \texttt{config\_v1.json}:

\textbf{Optimizer (AdamW):}
\begin{itemize}
    \item $\beta_1$ (\texttt{adam\_b1}): 0.8
    \item $\beta_2$ (\texttt{adam\_b2}): 0.99
    \item Learning rate (\texttt{learning\_rate}): 0.0002 ($2 \times 10^{-4}$)
    \item Learning rate decay (\texttt{lr\_decay}): 0.999 (exponential decay per epoch)
\end{itemize}

\textbf{Training Parameters:}
\begin{itemize}
    \item Batch size (\texttt{batch\_size}): 16
    \item Random seed (\texttt{seed}): 1234
    \item Number of workers (\texttt{num\_workers}): 4
    \item Training epochs:
    \begin{itemize}
        \item Comparative experiments (Section 4): 500 epochs for all models
        \item Final BemaGANv2 model: 3,100 epochs (set via \texttt{--training\_epochs} in \texttt{train.py})
    \end{itemize}
\end{itemize}

\textbf{Loss Function Weights:}
\begin{itemize}
    \item Feature matching loss weight ($\lambda_{fm}$): 2 (implemented in \texttt{models.py} as \texttt{return loss*2})
    \item Mel-spectrogram loss weight ($\lambda_{mel}$): 45 (implemented in \texttt{train.py} as \texttt{F.l1\_loss(y\_mel, y\_g\_hat\_mel) * 45})
\end{itemize}

\subsection{Audio Processing Configuration}

Configuration parameters from \texttt{config\_v1.json}:

\textbf{Mel-Spectrogram Parameters:}
\begin{itemize}
    \item Number of mel bands (\texttt{num\_mels}): 80
    \item FFT size (\texttt{n\_fft}): 1024
    \item Window size (\texttt{win\_size}): 1024
    \item Hop size (\texttt{hop\_size}): 256
    \item Sampling rate (\texttt{sampling\_rate}): 24,000 Hz
    \item Frequency range: \texttt{fmin} = 0 Hz, \texttt{fmax} = 12,000 Hz
    \item Segment size (\texttt{segment\_size}): 8,192 samples
\end{itemize}

\textbf{Generator Configuration:}
\begin{itemize}
    \item Activation (\texttt{activation}): ``snakebeta''
    \item Snake log scale (\texttt{snake\_logscale}): true
    \item Upsample rates (\texttt{upsample\_rates}): [8, 8, 2, 2]
    \item Upsample kernel sizes (\texttt{upsample\_kernel\_sizes}): [16, 16, 4, 4]
    \item Initial channel (\texttt{upsample\_initial\_channel}): 512
    \item Resblock type (\texttt{resblock}): ``1''
    \item Resblock kernel sizes (\texttt{resblock\_kernel\_sizes}): [3, 7, 11]
    \item Resblock dilation sizes (\texttt{resblock\_dilation\_sizes}): [[1,3,5], [1,3,5], [1,3,5]]
\end{itemize}

\textbf{Discriminator Configuration:}

\textit{Multi-Resolution Discriminator (MRD):}
\begin{itemize}
    \item Resolutions (\texttt{resolutions}): [[1024, 120, 600], [2048, 240, 1200], [512, 50, 240]]
    \item Format: [n\_fft, hop\_length, win\_length]
\end{itemize}

\textit{Multi-Period Discriminator (MPD):}
\begin{itemize}
    \item Periods (\texttt{mpd\_reshapes}): [2, 3, 5, 7, 11]
\end{itemize}

\textit{Multi-Envelope Discriminator (MED):}
\begin{itemize}
    \item Five sub-discriminators with different envelope extraction modes (defined in \texttt{models.py}):
    \begin{itemize}
        \item $f = -1$: Lower envelope (negative of absolute Hilbert transform)
        \item $f = 0$: Original signal (no filtering)
        \item $f = 1$: Upper envelope (absolute value of Hilbert transform)
        \item $f = 300$: Envelope after 300 Hz low-pass Butterworth filter
        \item $f = 500$: Envelope after 500 Hz low-pass Butterworth filter
    \end{itemize}
\end{itemize}

\textbf{Other Settings:}
\begin{itemize}
    \item Use spectral norm (\texttt{use\_spectral\_norm}): false
    \item Discriminator channel multiplier (\texttt{discriminator\_channel\_mult}): 1
\end{itemize}

\subsection{Dataset Preparation}

\textbf{LJSpeech Dataset:}
\begin{itemize}
    \item Source:\url{https://keithito.com/LJ-Speech-Dataset/}
    \item Original format: 22,050 Hz, 16-bit PCM WAV files
\end{itemize}

\textbf{Resampling Process:}

As described in README.md, the resampling procedure is:

\begin{enumerate}
    \item Download LJSpeech dataset
    \item Move all WAV files to \texttt{LJSpeech-1.1/wavs0}
    \item Run the following script:
\end{enumerate}
\begin{lstlisting}[language=Python]
import os
import glob

input_dir = "/Your_path/LJSpeech-1.1/wavs0"
output_dir = "/Your_path/LJSpeech-1.1/wavs"
os.makedirs(output_dir, exist_ok=True)

for file in glob.glob(f"{input_dir}/*.wav"):
    filename = os.path.basename(file)
    out_path = os.path.join(output_dir, filename)
    !ffmpeg -y -loglevel panic -i "{file}" -ar 24000 "{out_path}"
\end{verbatim}

This resamples all audio files to 24,000 Hz.

\textbf{Runtime Audio Normalization:}

The normalization is performed in \texttt{meldataset.py} during data loading:

\begin{verbatim}
audio, sampling_rate = load_wav(filename)
audio = audio / MAX_WAV_VALUE  # MAX_WAV_VALUE = 32768.0
if not self.fine_tuning:
    audio = normalize(audio) * 0.95  # librosa normalize 
                                      # then scale by 0.95
\end{lstlisting}

\textbf{Training/Validation Split:}

Files are specified in \texttt{train.py} arguments:
\begin{itemize}
    \item Training file: \texttt{--input\_training\_file} (default: \texttt{LJSpeech-1.1/training.txt})
    \item Validation file: \texttt{--input\_validation\_file} (default: \texttt{LJSpeech-1.1/validation.txt})
\end{itemize}

The \texttt{get\_dataset\_filelist()} function in \texttt{meldataset.py} reads these files.

\textbf{Data Augmentation:}

In \texttt{meldataset.py}:
\begin{itemize}
    \item Random seed: 1234 (\texttt{random.seed(1234)})
    \item Random shuffling applied if \texttt{shuffle=True}
    \item Random segment extraction: 8,192 samples cropped randomly during training
    \item Zero-padding applied if audio is shorter than segment size
\end{itemize}

\subsection{DTraining Procedure}

\textbf{Training Script:} \texttt{train.py}

\textbf{Key Training Steps:}

\begin{enumerate}
    \item \textbf{Initialization:}
    \begin{itemize}
        \item Random seed set via \texttt{torch.manual\_seed(h.seed)} and \texttt{torch.cuda.manual\_seed(h.seed)}
        \item Generator and discriminators (MRD, MED) initialized
    \end{itemize}
    
    \item \textbf{Optimizer Setup:}
\begin{lstlisting}[language=Python]
optim_g = torch.optim.AdamW(generator.parameters(), 
                            h.learning_rate, 
                            betas=[h.adam_b1, h.adam_b2])
optim_d = torch.optim.AdamW(
    itertools.chain(med.parameters(), mrd.parameters()),
    h.learning_rate, betas=[h.adam_b1, h.adam_b2])
\end{lstlisting}
    
    \item \textbf{Learning Rate Scheduler:}
\begin{lstlisting}[language=Python]
scheduler_g = torch.optim.lr_scheduler.ExponentialLR(
    optim_g, gamma=h.lr_decay, last_epoch=last_epoch)
scheduler_d = torch.optim.lr_scheduler.ExponentialLR(
    optim_d, gamma=h.lr_decay, last_epoch=last_epoch)
\end{lstlisting}
    
    \item \textbf{Training Loop (per iteration):}
    \begin{itemize}
        \item Discriminator update:
        \begin{itemize}
            \item Forward pass with real and generated (detached) audio
            \item Compute \texttt{loss\_disc\_all = loss\_disc\_s + loss\_disc\_f} (MED + MRD)
            \item Backward pass and optimizer step
        \end{itemize}
        \item Generator update:
        \begin{itemize}
            \item Compute mel-spectrogram loss: \texttt{loss\_mel = F.l1\_loss(y\_mel, y\_g\_hat\_mel) * 45}
            \item Compute feature matching and adversarial losses
            \item Total loss: \texttt{loss\_gen\_all = loss\_gen\_s + loss\_gen\_f + loss\_fm\_s + loss\_fm\_f + loss\_mel}
            \item Backward pass and optimizer step
        \end{itemize}
    \end{itemize}
    
    \item \textbf{Per-epoch operations:}
    \begin{itemize}
        \item Learning rate decay: \texttt{scheduler\_g.step()} and \texttt{scheduler\_d.step()}
    \end{itemize}
\end{enumerate}

\textbf{Logging and Checkpointing:}

From \texttt{train.py} default arguments:
\begin{itemize}
    \item Stdout interval (\texttt{--stdout\_interval}): 5 iterations
    \item Checkpoint interval (\texttt{--checkpoint\_interval}): 1,000 iterations
    \item Summary interval (\texttt{--summary\_interval}): 100 iterations (TensorBoard)
    \item Validation interval (\texttt{--validation\_interval}): 1,000 iterations
\end{itemize}

\textbf{Checkpoint Saving:}
\begin{itemize}
    \item Generator: saved as \texttt{g\_\#\#\#\#\#\#\#\#} (where \texttt{\#\#\#\#\#\#\#\#} is step number)
    \item Discriminators: saved as \texttt{do\_\#\#\#\#\#\#\#\#} including MRD, MED, and optimizer states
\end{itemize}

\textbf{Training Outputs:}
\begin{itemize}
    \item Training logs saved to CSV: \texttt{data\_BemaGanv2.csv} (includes epoch, steps, losses, time/batch)
    \item TensorBoard logs saved to: \texttt{\{checkpoint\_path\}/logs}
\end{itemize}

\subsection{Subjective Evaluation Protocol}

\textbf{MOS/SMOS Evaluation:}
\begin{itemize}
    \item Number of participants: 137
    \item Confidence interval: 95\%
    \item Rating scale: 5-point Likert (1-5)
    \item Audio categories: Short (2-10 seconds) and Long (90 seconds)
\end{itemize}

\subsection{Baseline Model Training}

For the comparative evaluation in Section 4, all models (including 
BemaGANv2, HiFi-GAN, and BigVGAN) were trained with identical settings 
to ensure fair comparison:

\begin{itemize}
    \item Sampling rate: 24 kHz
    \item Training epochs: 500
    \item Batch size: 16
    \item Random seed: 1234
    \item Same preprocessing pipeline
\end{itemize}

\subsection{Code and Model Availability}

\textbf{GitHub Repository:} \url{https://github.com/dinhoitt/BemaGANv2}

\textbf{Repository Structure (from README.md):}
\begin{verbatim}
BemaGANv2/
├── train.py
├── inference.py
├── config_v1.json
├── requirements.txt
├── models.py
├── meldataset.py
├── cp_BemaGanv2_MED_MRD/
└── README.md
\end{verbatim}

\textbf{Training Command:}
\begin{verbatim}
python train.py --config config_v1.json
\end{verbatim}

\textbf{Inference Command:}
\begin{lstlisting}[language=Python]
python inference.py --checkpoint_file [generator checkpoint file path]
\end{lstlisting}

\subsection{Software Dependencies}

From README.md:
\begin{itemize}
    \item Python version: 3.9
\end{itemize}

Installation:
\begin{verbatim}
pip install -r requirements.txt
\end{verbatim}

\subsection{Reproducibility}

\textbf{Random Seed Control:}
\begin{itemize}
    \item Config file: \texttt{"seed": 1234}
    \item Applied in \texttt{train.py}: \texttt{torch.manual\_seed(h.seed)} and \texttt{torch.cuda.manual\_seed(h.seed)}
    \item Applied in \texttt{meldataset.py}: \texttt{random.seed(1234)}
\end{itemize}

\textbf{Deterministic Settings:}
\begin{itemize}
    \item \texttt{torch.backends.cudnn.benchmark = True} (set in \texttt{train.py})
\end{itemize}

\subsection{Training Cost and Efficiency} \label{D.8}

Table~\ref{tab:training_cost} summarizes the computational cost and model complexity of BemaGANv2. All training was performed on a single NVIDIA A100 GPU (40GB) via Google Colab. Training times for the comparative experiments (500 epochs) and the final BemaGANv2 model (3,100 epochs) are estimated based on measured wall-clock time for 300 epochs ($\sim$35.5 hours), assuming linear scaling.

We note that the discriminators (MED and MRD) are used only during training and are discarded at inference time. Therefore, the effective model size for deployment is 13.95M parameters (Generator only), making BemaGANv2 lightweight and practical for real-time applications.

\begin{table}[htbp]
\centering
\caption{Training cost and model specifications for BemaGANv2.}
\resizebox{\columnwidth}{!}{%
\begin{tabular}{l|c}
\hline
\textbf{Specification} & \textbf{Value} \\ \hline
\multicolumn{2}{l}{\textit{Model Parameters}} \\ \hline
\quad Generator (AMP + Snake) & 13.95M \\
\quad MED & 49.37M \\
\quad MRD & 0.28M \\
\quad \textbf{Total} & \textbf{63.61M} \\ \hline
\multicolumn{2}{l}{\textit{Training Cost}} \\ \hline
\quad Hardware & 1$\times$ NVIDIA A100 (40GB) \\
\quad Peak VRAM usage & $\sim$36 GB (batch size 16) \\
\quad Time per epoch & $\sim$7.1 min \\
\quad Comparative experiments (500 epochs) & $\sim$59 hours ($\sim$2.5 days) \\
\quad Final model (3,100 epochs) & $\sim$367 hours ($\sim$15.3 days) \\ \hline
\multicolumn{2}{l}{\textit{Inference Speed}} \\ \hline
\quad Real-Time Factor (RTF) & 0.0097 \\
\quad Speed & $\sim$103$\times$ faster than real-time \\
\quad Hardware & NVIDIA A100 \\ \hline
\end{tabular}%
}
\label{tab:training_cost}
\end{table}

The inference speed was measured by generating audio of varying durations (10, 20, 60, and 90 seconds) and averaging the Real-Time Factor (RTF) over five runs after GPU warm-up. BemaGANv2 achieves an average RTF of 0.0097, corresponding to approximately 103$\times$ faster than real-time synthesis on an NVIDIA A100 GPU, confirming its suitability for practical deployment in real-time audio generation pipelines.

All experiments can be reproduced using the provided code and configuration files.

\section{Multi-Seed Variability Analysis}
\label{E}

To assess the reproducibility and stability of our results given the stochastic nature of GAN training, we trained BemaGANv2 (MED+MRD) with four different random seeds under identical conditions (500 epochs, batch size 16, all other hyperparameters as specified in ~\ref{D}). Table~\ref{tab:multiseed_short} and Table~\ref{tab:multiseed_long} report the per-seed results along with the mean and standard deviation for short-term and long-term audio evaluation, respectively.

\begin{table}[htbp]
\centering
\caption{Multi-seed variability of BemaGANv2 for short-term audio. Results are reported for four random seeds with mean $\pm$ standard deviation.}
\resizebox{\columnwidth}{!}{%
\begin{tabular}{l|c|c|c|c|c|c}
\hline
 & \textbf{FAD $\downarrow$} & \textbf{SSIM $\uparrow$} & \textbf{PCC $\sim 1$} & \textbf{MCD $\downarrow$} & \textbf{M-STFT $\downarrow$} & \textbf{Periodicity $\downarrow$} \\ \hline
Seed 1 & 1.045 & 0.867 & 0.947 & 1.751 & 1.312 & 0.144 \\
Seed 2 & 1.136 & 0.875 & 0.945 & 1.968 & 1.310 & 0.182 \\
Seed 3 & 0.785 & 0.873 & 0.943 & 1.796 & 1.311 & 0.169 \\
Seed 4 & 0.911 & 0.849 & 0.949 & 1.652 & 1.262 & 0.161 \\ \hline
\textbf{Mean $\pm$ Std} & \textbf{0.969$\pm$0.133} & \textbf{0.866$\pm$0.010} & \textbf{0.946$\pm$0.002} & \textbf{1.792$\pm$0.115} & \textbf{1.299$\pm$0.021} & \textbf{0.164$\pm$0.014} \\ \hline
\end{tabular}%
}
\label{tab:multiseed_short}
\end{table}

\begin{table}[htbp]
\centering
\caption{Multi-seed variability of BemaGANv2 for long-term audio. Results are reported for four random seeds with mean $\pm$ standard deviation.}
\resizebox{\columnwidth}{!}{%
\begin{tabular}{l|c|c|c|c|c|c}
\hline
 & \textbf{FAD $\downarrow$} & \textbf{SSIM $\uparrow$} & \textbf{PCC $\sim 1$} & \textbf{MCD $\downarrow$} & \textbf{M-STFT $\downarrow$} & \textbf{Periodicity $\downarrow$} \\ \hline
Seed 1 & 2.510 & 0.774 & 0.946 & 1.789 & 1.483 & 0.125 \\
Seed 2 & 2.607 & 0.770 & 0.935 & 1.833 & 1.540 & 0.136 \\
Seed 3 & 1.865 & 0.776 & 0.946 & 1.861 & 1.472 & 0.123 \\
Seed 4 & 2.681 & 0.784 & 0.945 & 1.78 & 1.514 & 0.124 \\ \hline
\textbf{Mean $\pm$ Std} & \textbf{2.416$\pm$0.324} & \textbf{0.776$\pm$0.005} & \textbf{0.943$\pm$0.005} & \textbf{1.820$\pm$0.029} & \textbf{1.502$\pm$0.027} & \textbf{0.127$\pm$0.005} \\ \hline
\end{tabular}%
}
\label{tab:multiseed_long}
\end{table}

The results demonstrate that BemaGANv2 exhibits low variability across different random seeds. Most metrics show small standard deviations---for example, PCC varies by only $\pm$0.002 in short-term and $\pm$0.005 in long-term evaluation, and SSIM varies by $\pm$0.010 and $\pm$0.005, respectively. M-STFT and Periodicity also show consistent performance with standard deviations below 0.03. FAD exhibits relatively higher variability (std of 0.133 for short-term and 0.324 for long-term), which is consistent with the known sensitivity of FAD to distributional estimation~\cite{b15, b41}. Overall, these results confirm that the performance advantages reported in the main text are robust and reproducible, and are not artifacts of a particular random initialization.

\section*{Conflict of interest}
The authors declare that there is no conflict of interest in this paper.

\section*{Declaration of Generative AI and AI-assisted Technologies in the Writing Process}

During the preparation of this work, the author(s) used ChatGPT in order to improve English translation and language clarity. After using this tool, the author(s) reviewed and edited the content and take full responsibility for it.



\bibliographystyle{elsarticle-num}

\end{document}